\documentclass[prb,superscriptaddress,twocolumn,floatfix,amsmath,amssymb]{revtex4}
\usepackage{graphicx}
\usepackage{dcolumn}
\usepackage{bm}
\usepackage{color}
\renewcommand{\vec}{\mathbf}

\begin{document}

\title{Resolution of two apparent paradoxes concerning quantum oscillations in underdoped high-$T_{c}$ superconductors}

\author{Xun Jia}
\affiliation{Department of Physics and Astronomy, University of
California Los Angeles\\ Los Angeles, CA 90095-1547}
\author{Pallab Goswami}
\affiliation{Department of Physics and Astronomy, Rice University, TX 777005}
\author{Sudip Chakravarty}
\affiliation{Department of Physics and Astronomy, University of
California Los Angeles\\ Los Angeles, CA 90095-1547}

\email{sudip@physics.ucla.edu}

\date{\today}

\begin{abstract}
Recent quantum oscillation experiments in underdoped high temperature superconductors seem to imply two paradoxes. 
The  first paradox concerns the apparent non-existence of the signature of the electron pockets in angle resolved photoemission spectroscopy (ARPES). The second paradox is a clear signature of a small electron pocket in quantum oscillation experiments, but no evidence as yet of the corresponding hole pockets of approximately double the frequency of the electron pocket. This hole pockets should be present  if the Fermi surface reconstruction is due to a commensurate density wave, assuming that Luttinger sum rule relating the area of the pockets and the total number of charge carriers holds. Here we provide  possible resolutions of these apparent paradoxes from the  commensurate $d$-density wave theory.  To address the first paradox we have computed the ARPES spectral function
 subject to correlated disorder, natural to a class of experiments relevant to the materials studied in quantum oscillations.
The intensity of the spectral function is significantly reduced for
the electron pockets for an intermediate range of disorder correlation
length, and typically less than half the hole pocket is visible, mimicking  Fermi arcs. Next we show from an exact transfer matrix calculation of the Shubnikov-de Haas oscillation that the usual disorder affects the electron pocket more significantly than the hole pocket. However, when, in addition, the  scattering  from vortices in the mixed state is included, it wipes out the frequency corresponding to the hole pocket. Thus, if we are correct, it will be necessary to do measurements at higher magnetic fields and even higher quality samples to recover the hole pocket frequency.

\end{abstract}


\maketitle
\section{Introduction}

High temperature superconductors have been addressed from a remarkable number of vantage
points. Nonetheless many of the basic questions still remain unresolved, and the notion of broken symmetries
arising from a Fermi liquid has been traditionally discarded in favor of many exotic ideas. Here we revisit
the Fermi liquid concept and a particular broken symmetry in response to a class of recent quantum oscillation experiments.~\cite{Doiron-Leyraud:2007,LeBoeuf:2007,Bangura:2008,Jaudet:2008,Yelland:2008,Sebastian:2008}
We and
others~\cite{Chakravarty:2008b,Jia:2008,Dimov:2008,Millis:2007,Podolsky:2008,Lin:2008,Morinari:2009}
have had some success in this respect. If these
theories and experiments are correct, one will have to radically
alter our twenty year-old view of these
superconductors.~\cite{Chakravarty:2008} The totality of phenomenology cannot of course be explained without serious
Fermi liquid corrections. But as long as the quasiparticle residue is finite, we hope that the low 
energy properties can be understood from our perspective.

Two paradoxes have arisen in the context of the quantum oscillation measurements. The first is the  contrast between Fermi arcs observed in angle
resolved photoemission (ARPES) experiments~\cite{Damascelli:2003} on
one hand and recent quantum oscillation experiments suggesting small
Fermi pockets in underdoped
$\mathrm{YBa_{2}Cu_{3}O_{6+\delta}}$ (YBCO),~\cite{Doiron-Leyraud:2007,LeBoeuf:2007,Bangura:2008,Jaudet:2008,Yelland:2008,Sebastian:2008}
on the other. This is particularly clear in recent ARPES experiments where an effort 
was made to examine YBCO with similar doping as in the quantum oscillation measurements.~\cite{Hossain:2008}

The second paradox is the non-existence of any evidence of the hole pockets in quantum oscillation measurements. Within the density wave scenario of wave vector ${\bf Q}=(\pi,\pi)$ (the lattice spacing set to unity), there should be two dominant frequencies in quantum oscillations. One corresponding  to the electron pocket at about 500T and the other corresponding to the hole pocket at around 900 T. These are of course rough numbers corresponding to approximately 10\% doping, assuming that the Luttinger sum rule is satisfied in the mixed state, that is, the quantum oscillations reflect the normal state even if the measurements may lie within the mixed state.

The concept of a broken symmetry is very powerful because deep
inside a phase a physically correct effective Hamiltonian can
address many important questions, whereas our inability to reliably
predict properties of even a single band Hubbard model, while widely
pursued, has been a limiting factor. This is not an empty exercise,
if new phenomena can be predicted or striking facts can be explained
with some degree of simplicity. That broken symmetry both dictates
and protects the nature of elementary excitations, determining
the properties of matter, is important to emphasize.

The suggested form of order, the $d$-density wave (DDW),~\cite{Chakravarty:2001} explains numerous
properties of these superconductors, including the concomitant
suppression of the superfluid density~\cite{Tewari:2001}, Hall
number,~\cite{Chakravarty:2002} and more recently the large enhancement of Nernst effect in the pseudogap state,~\cite{Tewari:2009}  in addition to the existence of a
single-particle gap of $d_{{x^2}-{y^2}}$ form above $T_c$. There are
also theoretical reasons why DDW is a possibility. It
competes favorably with other ordering tendencies in variational
studies of extended Hubbard models with nearest-neighbor repulsion
and pair-hopping terms.~\cite{Nayak:2000,Nayak:2002} It is also
realized in a class of two-leg ladder models with nearest-neighbor
repulsion.\cite{Schollwoeck:2003}  However, whatever form the
correct Hamiltonian takes, we know that it must favor $d$-wave
superconductivity (DSC). Such a Hamiltonian will almost certainly
favor DDW order as well, in light of the abundance of local
Hamiltonians which do not discriminate between DSC and DDW order. In
fact, two carefully designed difficult polarized neutron scattering
experiments have provided tantalizing direct evidence of DDW
order,~\cite{Mook:2002,Mook:2004} although other experiments have
claimed otherwise.~\cite{Stock:2002,Fauque:2006,footnote} 

A natural enemy of the pristine properties of matter is disorder
that is unavoidable in complex systems such as high temperature
superconductors. The role of disorder was emphasized in
the original proposal of DDW order as a
relevant competing order in the phase diagram of high-$T_{c}$
superconductors,~\cite{Chakravarty:2001} although our views of
disorder have greatly evolved during the intervening years. It is
this DDW order combined with disorder that would be the focus of the
present manuscript in resolving the paradoxes stated above. The disorder 
considered are of two different types: (1) scattering due to impurities and
defects and (2) scattering from vortices in the mixed states.

We consider two kinds of intrinsic disoredr: (a) Gaussian white noise and (b) correlated disorder with a finite correlation length. In
the momentum  space the  scattering rate for correlated disorder will decay as 
$\exp(-q^2l_D^2)$, where $\vec{q}$ is the momentum transfer
between the initial and the final states, $l_D$ being the correlation length. Therefore, because of its smaller size, the states corresponding to the electron pockets are scattered 
more than on the hole pockets. This is an interpretation of the phenomenon and is based on intrapocket scattering. An alternative interpretation involves the density of states on the Fermi surface of the electron pockets.   In contrast, for white noise, scattering is independent of momentum and affects both pockets similarly.  Disorder naturally  has a strong effect on
ARPES spectral function, which is sensitive to the coherence factors that are analogs of Wannier functions.  In Shubnikov-de Haas oscillations it is only the averaged effect of disorder that enters by 
determining the effective lifetime on the Fermi surface. Therefore, the role of disorder is quite different, as we shall explicitly see.

Correlated disorder is also
experimentally relevant. Unlike quantum oscillation experiments
which probes bulk properties, ARPES is inherently a surface probe. In the
relevant case of YBCO, as cleaved surfaces show that CuO and BaO
terminations give different contributions to the total photoemission
intensity, with a hole doping $n_{h}= 30\%$, almost irrespective of
the nominal bulk doping. This self-doping was controlled by
evaporating potassium in situ on the cleaved surface, so as to
reduce the hole content down to the value of underdoped bulk YBCO 
($\delta\approx 0.5$),~\cite{Hossain:2008} the doping level for
which many quantum oscillation experiments are carried out. The
potassium overlayer is likely to produce an effective correlated
disorder in the $\mathrm{CuO}$ plane.

To explain the second paradox we shall adapt an analysis of Stephen~\cite{Stephen:1992} to include a normal state that exhibits DDW order.  We shall see that the relativistic character of the nodal fermions of the hole pocket, as opposed to the nonrelativistic nature of the charge carriers of the electron pocket provides a possible explanation. If we denote the Dingle factor of the electron pocket by ${\cal D}_{e}=e^{-\pi/\omega_{c}\tau_{v}}$, the Dingle factor of the hole pocket is ${\cal D}_{h}\approx {\cal D}_{e}^{4.4}$. This huge suppression may be the resolution of the missing frequency from the hole pocket. Here, $\omega_{c}$ is the cyclotron frequency corresponding to the electron pocket and $1/\tau_{v}$ is the scattering rate of the electrons from the vortices in the mixed state. The analysis of Stephen also  leads to a  tiny shift of the relative frequency of the quantum  oscillations in the mixed state,  of the order of $10^{-6}$. Thus, there is enough leeway that even a very large error in this estimate   will not affect our conclusions.

The organization of the manuscript is as follows. In Sec. II we  compute
the ARPES spectral function and show that disorder can destroy the evidence
of electron pockets. Section III is devoted to an exact transfer matrix
computation of Shubnikov-de Haas oscillations and the effect of disorder
on it. Section IV contains a discussion of scattering of quasiparticles of the putative normal state from
the vortices in the mixed state. In Sec. V we briefly summarize the salient features of our work.

\section{Spectral Function}

\subsection{Hamiltonian}
The Hamiltonian of commensurate DDW order in terms of the fermion creation and destruction  operators, $c^{\dagger}_{\bf k}$ and $c_{\bf k}$, in the
momentum space is 
\begin{equation}
\begin{split}
    H_1 = &\sum_{\vec{k}\in RBZ}\left(\epsilon_\vec{k}c^\dag_\vec{k}c_\vec{k}
        +\epsilon_{\vec{k}+\vec{Q}}c^\dag_{\vec{k}+\vec{Q}}c_{\vec{k}+\vec{Q}}\right)\\
        &+\sum_{\vec{k}\in RBZ}
        (iW_\vec{k}c^\dag_\vec{k}c_{\vec{k}+\vec{Q}}+h.c.),
\end{split}
    \label{Eq:pureHamiltonian}
\end{equation}
where the ordering wave vector $\vec{Q} = (\pi, \pi)$, and
$\epsilon_\vec{k}$ is the single particle spectra. The lattice
constant is set to be unity for simplicity. The reduced Brillouin
zone (RBZ) is bounded by $k_y \pm k_x = \pm\pi$. We define 
$\epsilon_{\bf k}$ by
\begin{eqnarray}
\epsilon_\vec{k} = &-& 2t(\cos k_x + \cos k_y)+ 4t'\cos
k_x \cos k_y \nonumber \\ &-& 2t''(\cos 2k_x + \cos 2k_y)
\end{eqnarray}
and the DDW gap  by
\begin{equation}
W_\vec{k} = \frac{W_0}{2}(\cos k_x - \cos k_y).
\end{equation}

\subsection{Disorder}
Potential disorder in real space
with a finite correlation length $l_{D}$ is modeled by
\begin{equation}
    \label{Eq:disorderRealSpace}
    V(\vec{r}) = \frac{g_V}{2\pi l_D^2}\int
    \mathrm{d}\vec{x}~\mathrm{e}^{-\frac{|\vec{r}-\vec{x}|^2}{2l_D^2}}G(\vec{x}),
\end{equation}
where the disorder averages are 
 $\langle G(\vec{x})\rangle=0$ and $\langle
G(\vec{x})G(\vec{y})\rangle=\delta(\vec{x}-\vec{y})$; the disorder intensity
is set by $g_V$. This disorder Hamiltonian in the momentum space is then 
\begin{equation}
    \label{Eq:disHamtKSpace}
    H_2 = \sum_{\vec{k}_1,\vec{k}_2\in
    BZ}V(\vec{k}_1,\vec{k}_2)c^\dag_{\vec{k}_1}c_{\vec{k}_2}+h.c.,
\end{equation}
where the matrix elements are
\begin{equation}
    \label{Eq:disHamtelement}
    V(\vec{k},\vec{k}+\vec{q})=\frac{g_V}{2\pi} e^{-\frac{q^2l_D^2}{2}}u(\vec{q}),
\end{equation}
and $u(\vec{q})$ is 
\begin{equation}
    u(\vec{q})=\frac{1}{2\pi}\int\mathrm{d}\vec{y}~G(\vec{y})
    \mathrm{e}^{-i\vec{q}\cdot\vec{y}},
\end{equation}
satisfying the
conditions of $\langle u(\vec{q})\rangle=0$ and $\langle
u(\vec{q})u(\vec{q'})\rangle= \delta(\vec{q}+\vec{q}')$. In practice, we
generate $u(\vec{q})$ directly with the desired statistical properties
and then compute the matrix elements in Eq.~(\ref{Eq:disHamtKSpace}).

\subsection{Computation of ARPES spectral function}
Once the full Hamiltonian $H=H_1+H_2$ is generated, it is
diagonalized  by the transformation
$c_\vec{k}=\sum_{l}P_{\vec{k},l}\gamma_l$, where $\gamma_l$ is the
annihilation operator of quasiparticles with energy $E_l$. The
coefficients $P_{\vec{k},l}$ and energy $E_l$ are obtained through
an exact numerical diagonalization procedure. Finally, the ARPES spectral
function $A(\vec{k},\omega)$ at a temperature $T$ is given by:
\begin{equation}
    \label{Eq:Akomega}
    A(\vec{k},\omega)=2\pi\sum_l|P_{{\bf k}l}|^2n_l\delta(\omega-E_l),
\end{equation}
where $n_l=1/[1+\exp((E_l-\mu)/k_BT)]$ is the fermion occupation
number. Note that the numerical implementation of
Eq.~(\ref{Eq:Akomega}) requires an approximation of the delta
function by, for example, a Lorentzian distribution.

\begin{figure}[htb]
    \centering
  \includegraphics[width=8.5 cm]{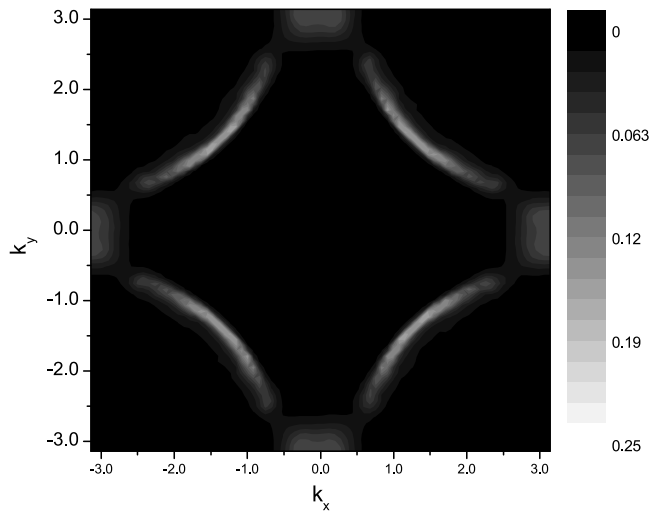}\\
  \caption{ The spectral function $A(\vec{k},\omega)$ at $\omega=\mu$
 with correlated disorder corresponding to $l_D=4$.
  The remaining parameters are stated in the text.}
  \label{Fig:SpectralFunction}
\end{figure}

We discretize the BZ with a mesh of size $80\times 80$, and
diagonalize the corresponding Hamiltonian. The parameters we choose
for YBCO at $10\%$ doping are: $t=0.3~\mathrm{eV}$, $t'=0.3t$,
$t''=t'/9.0$, and $W_0=0.0825~\mathrm{eV}$, same as
before.~\cite{Chakravarty:2008b} The chemical potential $\mu$ is set
to be $-0.2627~\mathrm{eV}$. These parameters yield a hole doping of
$n_h\sim 10\%$. The temperature $T=10~\mathrm{K}$ is chosen, where a
typical ARPES experiment is performed.  For  $g_V/(2\pi) = 0.1t$, the  quasiparticle life time for $l_{D}=0$
is of the order $\tau\sim 10^{-12}~\mathrm{s}$  from Fermi's golden
rule, which is a reasonable value.~\cite{Bangura:2008,Sebastian:2008} Note that the band width is $8
t$. The scattering rate for correlated disorder for a finite $l_{D}\sim 4$ is considerably smaller, as can be seen from  Eq.~\ref{Eq:disHamtelement}. The final spectral function is obtained by averaging over $20$ -
$50$ disorder configurations until no difference is detected upon
further averaging. A typical result with disorder correlation length
$l_D = 4$, in units of the lattice constant, is plotted in
Fig.~\ref{Fig:SpectralFunction}. The electron pockets are barely
visible, resembling experimental observations. From Fermi's golden
rule, the scattering rate is proportional to the square  of the
matrix element between the initial and the final states,
$V(\vec{k},\vec{k}+\vec{q})\sim \exp(-q^2l_D^2/2)$ (see
Eq.~(\ref{Eq:disHamtelement})), where $\vec{q}$ is the momentum
transfer in the scattering process. On average, the scattering rate
is therefore proportional to $\exp(-q_t^2l_D^2/2)$, where $q_t$ is a
typical momentum transfer. In particular, $q_t$ is roughly the size
of a pocket. Because  electron pockets are much smaller than  hole
pockets, the scattering rate is greater for electron pockets.
The hole
pockets centered at  $(\pm\pi/2,\pm\pi/2)$ have vanishingly small
spectral function on the back side due to the coherence factors and
appear as Fermi arcs instead,~\cite{Chakravarty:2003} whose lengths are further reduced by disorder. 

\begin{figure}[htb]
    \centering
  \includegraphics[width=8.5 cm]{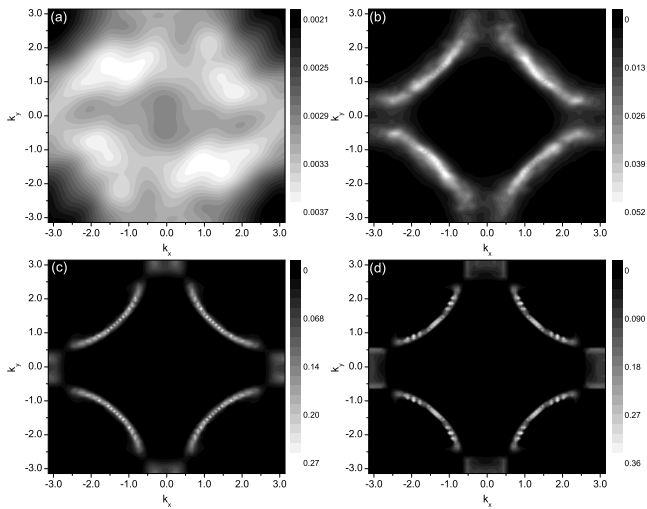}\\
  \caption{The spectral function $A(\vec{k},\omega=\mu)$
  for $g_V/(2\pi)=0.1t$. The correlation lengths are $l_D=0,2,8,16$
  from panels (a) through (d), respectively. The remaining parameters are given in the text.}
  \label{Fig:SFonCorrelation}
\end{figure}

We also demonstrate the dependence of the spectral function
$A(\vec{k},\omega=\mu)$ on $l_D$ in Fig.~\ref{Fig:SFonCorrelation}.
The disorder correlation lengths are $l_D=0,2,8,16$ for panels (a) through
(d), respectively. There are three distinct
regimes depending on $l_D$. For small $l_{D}$, the electron and hole
pockets will be almost equally scattered. As a consequence, the
spectral function is smeared out everywhere; see
Fig.~\ref{Fig:SFonCorrelation}(a). For intermediate values of
$l_{D}$, for example $l_D=2$ in Fig.~\ref{Fig:SFonCorrelation}(b)
and $l_D=4$ in Fig.~\ref{Fig:SpectralFunction}, scattering is
more prominent for electron pockets, resulting in a picture
consisting of only four Fermi arcs. Finally, as the correlation
length $l_D$ increases further, Fig.~\ref{Fig:SFonCorrelation}(c)
and (d), the electron pockets reappear. Indeed, though more
disorder scattering occurs on the electron pockets, the spatial
variation of disorder, hence the net effect of disorder, becomes
weaker.

\begin{figure}[htb]
    \centering
  \includegraphics[width=7 cm]{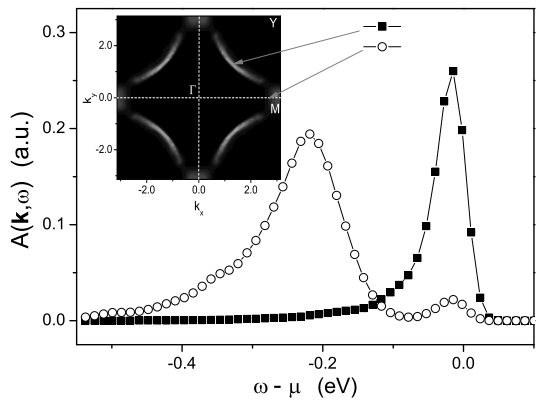}\\
  \caption{The energy dependence of the spectral function
  $A(\vec{k},\omega)$ at two $\vec{k}$ points in the Brillouin zone, as indicated in the insert.}
  \label{Fg:SFatGivenK}
\end{figure}

To characterize the energy dependence of the spectral function, we
compute $A(\vec{k},\omega)$ as a function of $\omega$ at two
$\vec{k}$ points in the Brillouin zone. One of them is at the intersection
 $\Gamma Y$ line with the inner side of the Fermi surface, the other is situated on
the electron pocket along the direction $\Gamma M$;  see
Fig.~\ref{Fg:SFatGivenK}. The disorder correlation length was chosen to be 
$l_D=4$. Although there are peaks at
$\omega\sim\mu$ for both,  the peak corresponding to the electron pocket is significantly suppressed by disorder; the second peak  at $\omega-\mu\sim -0.2\mathrm{eV}$ is clearly an artifact of our simple theory and such high energy states would surely decay once correlation effects are taken into account by the creation of particle-hole pairs.

\begin{figure}[htb]
    \centering
  \includegraphics[width=8 cm]{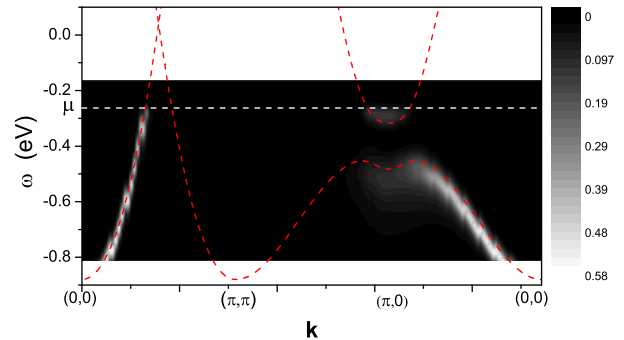}\\
  \caption{(Color online)The gray scale plot of the spectral function
  $A(\vec{k},\omega)$ as a function of both $\vec{k}$ and $\omega$. Red dashed
  curves
  indicate the quasi-particle dispersion relation. A horizontal dashed line shows
  the  chemical potential $\mu$. The remaining parameters are given in the text.}
  \label{Fg:SFkomega}
\end{figure}

In Fig.~\ref{Fg:SFkomega},  $A(\vec{k},\omega)$ is plotted  
as a function of both $\vec{k}$ and $\omega$. The horizontal axis
is along the path $\Gamma\to Y \to M\to\Gamma$ in the
Brillouin zone. The spectral
function is negligibly small outside the reduced Brillouin zone bounded by
$k_x\pm k_y=\pm\pi$,~\cite{Chakravarty:2003} and consequently
there are no peaks in the central region of Fig.~\ref{Fg:SFkomega}.
Close to $\vec{k}=(\pi,0)$ and $\omega=\mu$,
$A(\vec{k},\omega)$ has very small intensity due to long range
correlated disorder, consistent with our previous observation that the electron
pockets are most likely unobservable.

\section{Shubnikov-de Haas oscillations}
\subsection{The transfer matrix method}
Let us now consider the effect of disorder on Shubnikov-de Haas
(SdH)  oscillations of the conductivity, $\sigma_{xx}$. The
tight-binding Hamiltonian on a square lattice in a sample of
dimension $N\times M$, with the lattice constant set to unity, is
\begin{equation}\label{Eq:HamiltRealspace}
    H=\sum_{\vec{i}}\epsilon_\vec{i}c_\vec{i}^\dag c_\vec{i}+\sum_{
    \vec{i},\vec{j}}t_{\vec{i},\vec{j}}~\mathrm{e}^{ia_{\vec{i},\vec{j}}}c_\vec{i}^\dag
    c_\vec{j}+h.c.,
\end{equation}
where $c_{\vec{i}}$ is the fermionic annihilation operator at the site
$\vec{i}$. The spin degrees of freedom are omitted for simplicity.
The hopping amplitude $t_{\vec{i},\vec{j}}$ vanishes except for
nearest and next nearest neighbors. To include
two-fold commensurate DDW order, the nearest neighbor
hopping amplitudes are chosen to be
\begin{equation}\label{Eq:DDWhopping}
\begin{split}
    t_{\vec{i},\vec{i}+\hat{\vec{x}}}&=-t+\frac{iW_0}{4}(-1)^{(n+m)},\\
    t_{\vec{i},\vec{i}+\hat{\vec{y}}}&=-t-\frac{iW_0}{4}(-1)^{(n+m)},\\
\end{split}
\end{equation}
where $(n,m)$ are a pair of integers labeling a site:
$\vec{i}=n\hat{\vec{x}}+m\hat{\vec{y}}$, and $W_0$ is the DDW gap;
for the next nearest hopping $t_{\vec{i},\vec{j}}=t'$. The on-site
impurity energy $\epsilon_\vec{i}$ is defined by
\begin{equation}
    \label{Eq:disorderdiscrete}
    \epsilon_\vec{i}=\frac{V_0}{Z}\sum_\vec{r}G_\vec{r}\mathrm{e}^{-\frac{|\vec{r}-\vec{i}|^2}{2l_D^2}},
\end{equation}
which is analogous to Eq.~(\ref{Eq:disorderRealSpace}). To model
correlated disorder, we set the disorder averages  $\langle G_\vec{r}\rangle=0$ and $\langle
G_\vec{r}G_{\vec{r}'}\rangle=\delta_{\vec{r},\vec{r}'}$, and
$Z=\sum_\vec{r}e^{-|\vec{r}|^2/2l_D^2}$ is a
normalization factor. $V_0$ parameterizes the disorder intensity.
Note that Eq.(\ref{Eq:disorderdiscrete}) reduces to
$\epsilon_\vec{i}=V_0G_\vec{i}$ in the limit $l_D\to 0$, and
$\epsilon_\vec{i}$ becomes uncorrelated random variables. A
constant perpendicular magnetic field $B$ is included via the
Peierls phase factor $a_{\vec{i},\vec{j}}=\frac{2\mathrm{\pi}
e}{h}\int_\vec{j}^\vec{i}\vec{A}\cdot\mathrm{d}\vec{l}$, where
$\vec{A}=(0,-Bx,0)$ is the vector potential in the Landau gauge. We note
that a perpendicular magnetic field even as large as 60 T has little
effect on DDW order.~\cite{Nguyen:2002}

In this section we choose $t=0.29~\mathrm{eV}$,
$t'=0.1~\mathrm{eV}$, and $W_0=0.065~\mathrm{eV}$. The chemical
potential is set to be $\mu=-0.28~\mathrm{eV}$. Note that these
parameters are slightly different from those in the previous section, although the hole doping is again $\sim
10\%$. We have left out the  third nearest neighbor hopping, which
greatly complicates the transfer matrix calculation without offering
any particular insight. The disorder intensity $V_{0} = 0.4
t$ leads to a quasi-particle life time of the order  of $\sim
10^{-12}~\mathrm{s}$ in the limit of $l_D=0$. The magnetic field
ranges from $B=20~\mathrm{T}$ to $B=75~\mathrm{T}$, representative
of the quantum oscillation experiments. The only relevant length
scale here is the magnetic length $l_B=\sqrt{\hbar/eB}$, which for
 $B=20~\mathrm{T}$ is $\sim 15 a$, $a$ being lattice constant equal to
$3.85\textrm{\AA}$.

Now consider a quasi-1D system, $N\gg M$, with a periodic boundary
condition along y-direction. Let $\Psi_n = (\psi_{n,1},\psi_{n,2},
\ldots, \psi_{n,M})^T$ be the amplitudes on the slice $n$ for an
eigenstate with a given energy $E$, then the amplitudes on three
successive slices satisfy the relation
\begin{equation}
    \label{Eq:transfermatrix}
    \left[
      \begin{array}{c}
        \Psi_{n+1} \\
        \Psi_{n} \\
      \end{array}
    \right] = \left[
                \begin{array}{cc}
                  T_n^{-1}(E-H_n) & -T_n^{-1}T_{n-1} \\
                  1 & 0 \\
                \end{array}
              \right]\left[
                       \begin{array}{c}
                         \Psi_n \\
                         \Psi_{n-1} \\
                       \end{array}
                     \right],
\end{equation}
where $H_n$ is the Hamiltonian within the slice $n$, and the matrix
$T_n$ corresponds to the hopping between the slices $n$ and $n+1$.
$T_n$ is tridiagonal, as electrons can hop from a site on slice
$n$ to three sites on the slice $n+1$. All postive Lyapunov exponents
of the transfer matrix,~\cite{Kramer:1996}
$\gamma_1>\gamma_2>\ldots>\gamma_{M}$, are computed by iterating
Eq.~(\ref{Eq:transfermatrix}) and performing orthonormalization
regularly. The convergence of this algorithm is guaranteed by the
well known Oseledec theorem.~\cite{Oseledec:1968}
For the above parameters a
transverse dimension corresponding to $M=40$ is sufficient. Equation~(\ref{Eq:transfermatrix}) was iterated $10^5$ to $10^6$ times until the relative errors of less than $1\%$ of all the
Lyapunov exponents were achieved.

\subsection{Computation of $\sigma_{xx}$}
The conductivity
$\sigma_{xx}$ at zero temperature is obtained from the Landauer
formula:~\cite{Fisher:1981,Baranger:1989,Kramer:1993}
\begin{equation}\label{Eq:landauer}
    \sigma_{xx}(B)=\frac{e^2}{h}\sum_{i=1}^{M} \frac{1}{\cosh^2
    (M\gamma_i)}.
\end{equation}
\begin{figure}[htb]
    \centering
  \includegraphics[width=6.5 cm]{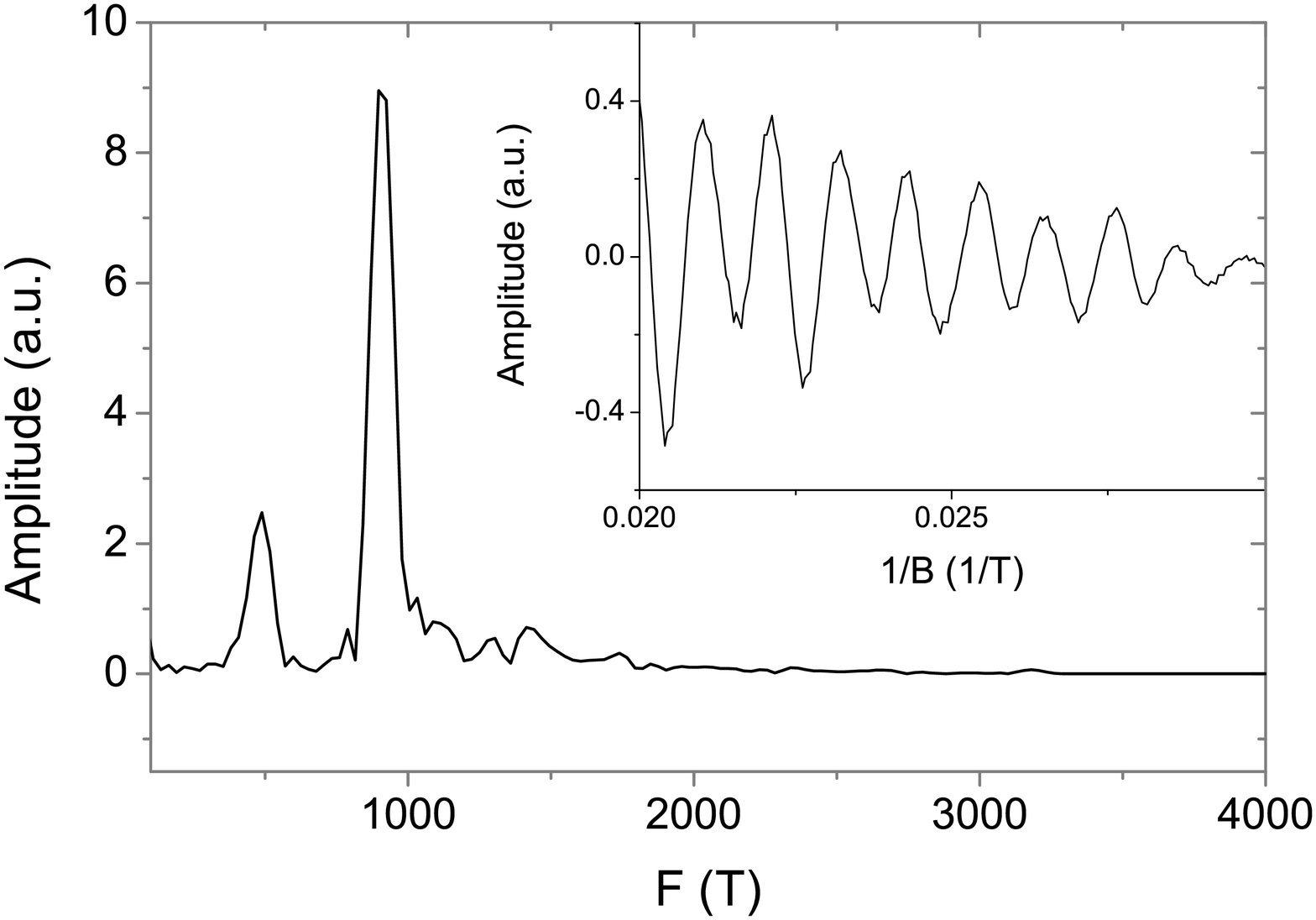}\\
  \caption{Fourier transform of SdH oscillations at zero temperature
  for $l_D=0$ in arbitrary units. Insert shows oscillations as a function of the
  inverse magnetic field.  Note the presence of higher harmonics. The parameters are described in the text.}
  \label{Fg:SdHzeroT}
\end{figure}
The SdH oscillations of $\sigma_{xx}$ at zero temperature for
 $l_D=0$ is shown in the insert of Fig.~\ref{Fg:SdHzeroT}. A third
order polynomial was used to subtract the background. The
Fourier transform is shown in the main panel of
Fig.~\ref{Fg:SdHzeroT} . Clearly, there are two main oscillation
frequencies $F_1=490\pm 30 \mathrm{T}$ and $F_2=915\pm 40
\mathrm{T}$, corresponding to electron and hole pockets
respectively, although a few harmonics are also visible. Though the
first peak at $F_1$ agrees with experimental observations, the
second peak at $F_2$ has not yet been observed. 
\begin{figure}[htb]
    \centering
  \includegraphics[width=6.5 cm]{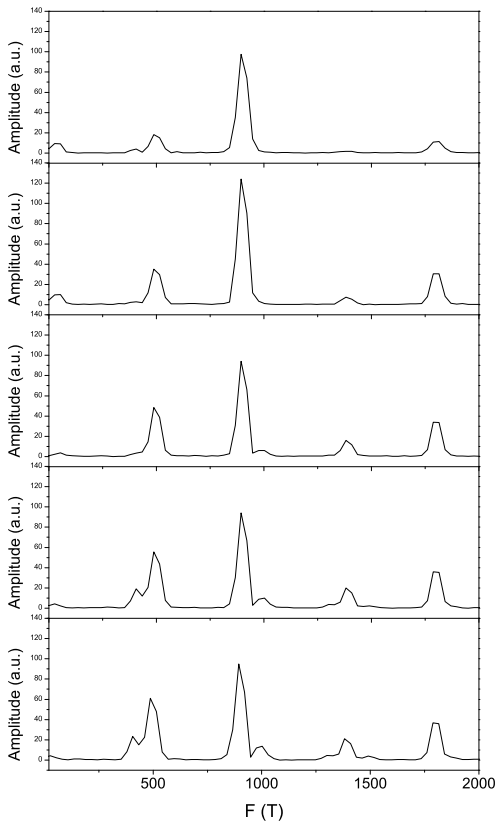}\\
  \caption{Fourier transform of SdH oscillations for correlated disorder
    at zero temperature. The correlation length $l_D=1,2,4,8,16$
    for panels (a) through (e), respectively. The remaining parameters are described in the text.}
  \label{Fg:SdHzeroT_LD}
\end{figure}

For correlated disorder, the Fourier transform of
$\sigma_{xx}$ is plotted in Fig.~\ref{Fg:SdHzeroT_LD}.
In
all the cases, two main frequencies of $F_1\sim
500\mathrm{T}$ and $F_2 \sim 900 \mathrm{T}$ are prominent. As for $A({\bf k},\omega)$, an increase 
of  $l_D$ increases the amplitudes
of $F_1$ and $F_2$, because the effective disorder
becomes weaker with the increased correlation length.
Also the higher harmonics are more visible. There is, however, a sharp distinction
between the dependence of the two physical quantities, which becomes  clear when 
we consider the white noise case:  $A({\bf k},\omega)$ is completely smeared out in
the momentum space because white noise scatters between all possible wave vectors; see Fig.~\ref{Fig:SFonCorrelation} (a). The coherence factors are of crucial importance for the spectral function.
In contrast, the SdH oscillations are damped by the Dingle factor, which is parametrized by a single
lifetime and disorder enters in an averaged sense. This striking contrast is clear in 
Fig.~\ref{Fg:SdHzeroT}. The surprise is that impurity scattering affects the electron pocket more than the hole pocket, which is remarkably robust even for white noise. There must be a piece of physics missing, if we are to explain why the hole pocket is not observed in SdH. This missing physics we argue is the vortex scattering rate which affects the two pockets very differently.

\section{Vortex scattering rate in the mixed state}

We have shown previously~\cite{Dimov:2008}  that the scattering rate of the DDW quasiparticles, corresponding to the electron pockets, from vortices  in the mixed state  is  given by,
\begin{equation}
\left(\frac{1}{\omega_{c}\tau_{v}}\right)_{e}\approx \frac{\Delta^{2}}{\hbar} \sqrt{\frac{\pi}{|\mu| \hbar \omega_c^{3}}},
\end{equation}
where the cyclotron frequency $\omega_{c}$ is determined from the band mass $m_{b}$ to be (restoring the lattice spacing $a$ here)
\begin{equation}
\frac{\hbar^{2}}{2m_{b}}\approx \left(2t'+4t''-\frac{W_{0}}{4}\right) a^{2}.
\label{eq:emstar}
\end{equation}
Here $\Delta^{2}$, a measure of the amplitude of the superconducting gap, is an average over the disordered configuration of the vortices, which is likely to be insensitive to the symmetry of the order parameter. Similarly, we have shown that the vortex scattering rate for the nodal DDW quasiparticles  at the hole pocket is given by (noting a typographical error in Ref.~\onlinecite{Dimov:2008})
\begin{equation}
\left(\frac{1}{\tau_{v}}\right)_{h}\approx \frac{\Delta^{2}}{\hbar}\frac{\sqrt{2 \pi}}{\sqrt{|\mu| \hbar \omega_{c}^{*}}}.
\end{equation}
The relevance of nodal quasi particle dispersion is also pointed out in Ref.~\onlinecite{Morinar:2009}
We have introduced a characteristic mass and a frequency scale by (nodal quasiparticle are actually massless)
\begin{eqnarray}
m^{*}&=&\frac{|\mu|}{v_Fv_D} ,
\label{eq:mstar}\\
\omega_{c}^{*}&=&\frac{eB}{m^{*}c}.
\end{eqnarray}
The reason for this is that these are precisely the scales that enter a SdH calculation of the nodal quasiparticles.~\cite{Goswami:2008} 
It can be seen from the above equations that  the vortex scattering rate for the nodal particles is proportional to $1/\sqrt{W_0}$. This idiverges  as $W_0\to 0$. But this is not unphysical, as, in that limit, there is a phase transition, where the Fermi surface reconnects.~\cite{Chakravarty:2001} The nodal fermions are obviously no longer a valid description.   However, our mean field theory cannot be trusted to predict the precise scattering rate at this quantum critical point. We have restricted ourselves to a regime that  we believe is deep inside the broken symmetry phase where our treatment should be a good guide, and where we believe that the present experiments are performed. A proper treatment of this scattering rate at the point where $W_0$ collapses is an open question that may shed light to the physics of the striking strange metal phase. 

The physical picture underlying the calculation of Stephen~\cite{Stephen:1992} is that the lifetime obtained from the imaginary part of the self-energy corresponds to a situation  as if the vortices are static impurities. Here $v_{F}=2\sqrt{2}at/\hbar$ is the magnitude of the velocity in a direction normal to the hole pocket and $v_{D}=W_{0}a/\sqrt{2}\hbar$ is the velocity in a direction orthogonal to it; neither $t'$ nor $t''$ enter in the leading order. Therefore,
\begin{equation}
\left(\frac{1}{\omega_{c}^{*}\tau_{v}}\right)_{h}\approx  \sqrt{2}\left(\frac{m^{*}}{m_{b}}\right)^{3/2}\left(\frac{1}{\omega_{c}\tau_{v}}\right)_{e}
\end{equation}
The cancellation of $\Delta^{2}$ is interesting but cannot of course be an exact result considering the approximations involved in Stephen's analysis.~\cite{Stephen:1992} The correctness of which relies entirely on the formulation of the averaged Green function of Stephen, where this average is carried out in the real space, and the nature of the normal state (whether or not it contains electron and hole pockets in the momentum space) does not enter, as the normal state Green function does not contain the superconducting order parameter. On the other hand, we had roughly estimated $m_{b}\approx 1.27 m_{e}$ and $m^{*}\approx 2.72 m_{e}$ for a typical set of parameters,~\cite{Dimov:2008} where $m_{e}$ is the free electron mass. Thus, it is reasonable that 
\begin{equation}
\left(\frac{1}{\omega_{c}^{*}\tau_{v}}\right)_{h}\approx 4.4\left(\frac{1}{\omega_{c}\tau_{v}}\right)_{e},
\label{eq:tauratio}
\end{equation}
which should be robust with respect small changes of the band parameters and the DDW gap. This would lead to a strong suppression of the  oscillations corresponding to the hole pocket as compared to the electron pocket, as the Dingle factors, ${\cal D}_{e,h}$ are exponentially sensitive to the product of the cyclotron frequency and the vortex scattering lifetime: ${\cal D}_{e}=e^{-\pi/(\omega_{c}\tau_{v})_{e}}$ and ${\cal D}_{h}={\cal D}_{e}^{4.4}$ . The previously estimated lifetime of the electrons,~\cite{Dimov:2008} 
\begin{equation}
\left(\frac{1}{\tau_{v}}\right)_{e}=1.5 \times 10^{12} \text{s}^{-1},
\label{eq:taue}
\end{equation}
for $B= 40\; \text{T}$ and $B_{c2}=60 \;\text{T}$ should be a rough guide.

The whole analysis is predicated on the assumption that the quantum oscillation frequencies are unshifted from the putative normal state (the DDW state in this case), for which we now provide some support from the analysis of Stephen;~\cite{Stephen:1992} see, however Ref.~\onlinecite{Chen:2009}. The formula for the relative frequency shift in terms of physical parameters (absolutely essential because they are all effective parameters) is  
\begin{equation}
\frac{\Delta F}{F} = \frac{\pi}{8}\frac{\Delta(H)^{4}}{\hbar\omega_{c}|\mu|^{3}},
\label{eq:shift}
\end{equation}
where $\Delta(H)$ is the zero temperature superconducting gap  in a magnetic field; for a $d$-wave superconductor we have to  use an appropriate   average, as above. 
The factor $\hbar\omega_{c}$ for free electrons is given
by
\begin{equation}
\hbar\omega_{c}=1.34\times 10^{-4} \times H (\textrm{Tesla}) {\mathrm eV}.
\end{equation}
As will be seen below, it will make little difference if we use the mass determined from experiments. The  magnetic field ranges  between $30 \textrm{T}$ and $65 \textrm{T}$. Let's take $H=40\; \textrm{T}$. Then
\begin{equation}
\hbar\omega_{c}=5.36 \times 10^{-3}  {\mathrm eV}.
\end{equation}
We know little about the zero temperature gap, especially in the underdoped regime, where there are fluctuation effects. As the simplest assumption, we use BCS mean field theory:
\begin{equation}
\Delta (H) = \Delta(0) \sqrt{1-H/H_{c2}}.
\end{equation}
and  $2\Delta(0) =3.52 \; \mathrm{T_{c}}$, which results in 
\begin{equation}
\Delta(0) = 8.6 \; \textrm{meV}
\end{equation}
for $T_{c}= 57.5\; {\mathrm K}$.~\cite{Doiron-Leyraud:2007,LeBoeuf:2007}
For simplicity we choose  $\Delta(0)\approx 10 \; \textrm{meV}$, and 
as a rough guide from experiments,~\cite{Doiron-Leyraud:2007,LeBoeuf:2007}  $H_{c2}\approx 60\; \textrm{T}$, although it will make little difference even  if it were 100 T.  
For $10\%$ doping we need a chemical potential $\mu \approx - 0.26\; \textrm{eV}$. Thus, we  get
\begin{equation}
\frac{\Delta F}{F} = 4.6 \times 10^{-6}.
\end{equation}
Even if this estimate were incorrect by several orders of magnitude, our conclusions that the frequency shift is negligible should be valid. 

\begin{figure}[htb]
    \centering
  \includegraphics[width=6.5 cm]{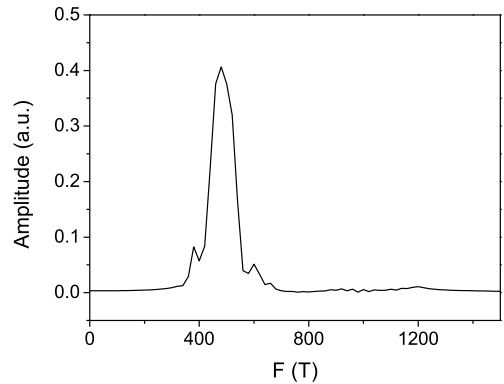}\\
  \caption{Fourier transform of SdH oscillations shown in Fig.~\ref{Fg:SdHzeroT}, after taking into account vortex scattering rates in the mixed state, as discussed in the text. Note that the amplitude of the peak corresponding to the hole pocket  at about 900 T is essentially wiped out. }
  \label{fig:Fig.7}
\end{figure}

Consider the white noise as an example; see Fig.~\ref{Fg:SdHzeroT}. We first filter out the two peaks, invert the Fourier transform, and then multiply by the Dingle factors corresponding to the vortex scattering rates discussed in this section. Using Eq.~\ref{eq:tauratio} and Eq.~\ref{eq:taue} and  Fourier transforming back this procedure essentially wipes out the peak corresponding to the frequency  of hole pocket in the resulting transform, as shown in Fig.~\ref{fig:Fig.7}. 
 
\section{Conclusions}

In summary, we have shown that two of the most puzzling features in
underdoped cuprates, the nonexistence of electron pockets  in ARPES and the lack evidence of the hole pockets in quantum oscillations can be plausibly  resolved by considering correlated disorder and the vortex scattering rates in the mixed state. With respect to the latter, it is crucial that the nodal fermions are described by a relativistic spectrum while the quasiparticles corresponding to the electron pocket are described by a non-relativistic spectrum determined by the bottom of the band.

It is important to determine if the calculation of Stephen can be so closely taken over after modifying for the presence of nodal fermions. We believe that it can be because the relative frequency shift turns out to be  very small, of the order of $10^{-6}$. Thus, even if the estimate is off by some orders of magnitude, we still have enough leeway. This argument has been recently challenged~\cite{Chen:2009} in a calculation of the density of states, however. Clearly, further investigations will be extremely valuable and are in progress.

One of the surprising conclusions of the present work, which we believe is on firm grounds, is that even very strong disorder, such as white noise, has only a modest effect on quantum oscillations, while it has a much larger effect on the ARPES spectra. This is because ARPES spectral function depends on the coherence factors, which act as Wannier functions that are naturally very sensitive to disorder. In quantum oscillations disorder appears primarily as an averaged lifetime in the Dingle factor. Elastic scattering has little direct effect on the Onsager quantization condition. 

It is also a firm conclusion of our work that elastic disorder scattering from impurities cannot be responsible for wiping out the hole pocket frequency, while keeping the electron pocket frequency more or less intact. In fact, quite the opposite seems to be true. Thus, the vortex physics in the mixed state appears to be of paramount importance, especially the contrast between the non-relativistic electrons around the anti-nodal  point and the relativistic nodal fermions at the nodal points of the Brillouin zone.

We also note an interesting paper~\cite{Stanescu:2008} regarding an analysis of various masses involved, which take into account electron-electron interactions. In the future, it would be interesting to pursue such an analysis of residual electron-electron interactions in the present context. 

\acknowledgments

This work is supported by NSF under Grant No. DMR-0705092. We thank Patrick Lee for many clarifying discussions.

\end{document}